\documentclass[utf8,12pt]{article}

\usepackage{graphicx}
\usepackage{hyperref}

\usepackage[onehalfspacing]{setspace}

\usepackage[english]{babel}

\usepackage{booktabs}

\usepackage{authblk}

\usepackage{amsmath}

\author[$\dagger$,$\star$]{Max Pellert}
\author[$\dagger$,$\star$]{Jana Lasser}
\author[$\dagger$,$\star$,$\ddag$]{Hannah Metzler}
\author[$\dagger$,$\star$]{David Garcia}

\affil[$\dagger$]{Complexity Science Hub Vienna, Vienna, Austria}
\affil[$\star$]{Section for Science of Complex Systems, Center for Medical Statistics, Informatics and Intelligent Systems, Medical University of Vienna, Vienna, Austria}
\affil[$\ddag$]{Institute for Globally Distributed Open Research and Education}

\title{Dashboard of sentiment in Austrian social media during COVID-19} 

\begin{document}

\maketitle

\begin{abstract}

To track online emotional expressions of the Austrian population close to real-time during the COVID-19 pandemic, we build a self-updating monitor of emotion dynamics using digital traces from three different data sources. This enables decision makers and the interested public to assess issues such as the attitude towards counter-measures taken during the pandemic and the possible emergence of a (mental) health crisis early on. We use web scraping and API access to retrieve data from the news platform derstandard.at, Twitter and a chat platform for students.  We document the technical details of our workflow in order to provide materials for other researchers interested in building a similar tool for different contexts. Automated text analysis allows us to highlight changes of language use during COVID-19 in comparison to a neutral baseline. We use special word clouds to visualize that overall difference. Longitudinally, our time series show spikes in anxiety that can be linked to several events and media reporting. Additionally, we find a marked decrease in anger. The changes last for remarkably long periods of time (up to 12 weeks). We discuss these and more patterns and connect them to the emergence of collective emotions. The interactive dashboard showcasing our data is available online under \href{http://www.mpellert.at/covid19_monitor_austria/}{http://www.mpellert.at/covid19\_monitor\_austria/}. Our work has attracted media attention and is part of an web archive of resources on COVID-19 collected by the Austrian National Library.

\end{abstract}

\section{Introduction}

In 2020, the outbreak of COVID-19 in Europe lead to a variety of counter-measures aiming to limit the spread of the disease. These include temporary lock downs, the closing of kindergartens, schools, shops and restaurants, the requirement to wear masks in public, and restrictions on personal contact. Health infrastructure was re-allocated with the goal of providing additional resources to tackle the emerging health crisis triggered by COVID-19. Such large-scale disruptions of private and public life can have tremendous influence on the emotional experiences of a population. 

Governments have to build on the compliance of their citizens with these measures. Forcing the population to comply by instituting harsh penalties is not sustainable in the longer run, especially in developed countries with established democratic institutions like in most of Europe. On the scale of whole nations, very strict policing also faces technical limits and diverts resources from other duties. In addition, recent research shows that, when compared to enforcement, the recommendation of measures can be a better motivator for compliance \cite{del2020differential}. Non-intrusive monitoring of emotional expressions of a population enables to identify problems early on, with the hope to provide the means to resolve them.

Due to the rapid development of the response to COVID-19, it is desirable to produce up-to-date observations of public sentiment towards the measures, but it is hard to quantify sentiment at large scales and high temporal resolution. Policy decisions are usually accompanied by representative surveys of public sentiment that, however, suffer from a number of shortcomings. First, surveys depend on explicit self-reports which do not necessarily align with actual behaviour \cite{baumeister2007psychology}. In addition, conducting surveys among larger numbers of people is time consuming and expensive. Lastly, a survey is always just a snapshot of public sentiment at a single point in time. Often, by the time a questionnaire is constructed and the survey has been conducted, circumstances have changed and the results of the survey are only partially valid.

Online communities are a complementary data source to surveys when studying current and constantly evolving events. Their digital traces reveal collective emotional dynamics almost in real-time. We gather these data in the form of text from platforms such as Twitter and news forums, where large groups of users discuss timely issues. We observe a lot of activity online, with clear increases during the nation-wide lock down of public life. For example, our data shows Austrian Twitter saw a 73\% increase in posts from 9 March 2020 compared to before (2019-01-01 until 2020-03-08). Livetickers at news platforms are a popular format that provides small pieces of very up-to-date news constantly over the course of a day. This triggers fast posting activity in the adjunct forum. By collecting these data in regular intervals, we face very little delay in data gathering and analysis and provide a complement to survey-based methods. Our setup has the advantage of bearing low cost while featuring a very large sample size. The disadvantages include more noise in the signal due to our use of automated text analysis methods, such as sentiment analysis. Additionally, if only information from one platform is considered, this might result in sampling a less representative part of the population than in surveys where participant demographics are controlled. However, systematic approaches to account for errors at different stages of research have been adapted to digital traces data \cite{senTotalErrorFramework2019}.

We showcase the monitoring of social media sentiment during the COVID-19 pandemic for the case of Austria. Austria is located in central Europe, serving as a small model for both Western Europe (especially Germany \cite{reutersAustriaKurzSays2020}) and Eastern Europe (e.g. Hungary \cite{hungarytodayCoronavirusOrbanGradual2020}). Therefore, the developments around COVID-19 in Austria have been closely watched by the rest of Europe. As the virus started spreading in Europe on a larger scale in February 2020, stringent measures were implemented comparatively early in Austria \cite{desvars2020interventionmeasures}. Using data from Austria allows us to build a quite extensive, longitudinal account of first hand discussions on COVID-19. Additionally, Austria's political system and its public health system have all the capacities of a developed nation to tackle a health crisis \cite{IntensivbettenOesterreich}. Therefore, we expect the population to express the personal, emotional reaction to the event, without being overwhelmed by lack of resources and resulting basic issues of survival.

Interactive online dashboards are an accessible way to summarize complex information to the public. During COVID-19, popular dashboards have conveyed information about the evolution of the number of COVID-19 cases in different regions of Austria \cite{austrianministryforhealthAmtlichesDashboardCOVID192020} and globally \cite{GlobalCoronaCases}. Other dashboards track valuable information such as world-wide COVID-19 registry studies \cite{thorlundRealtimeDashboardClinical2020}. Developers of dashboards include official governmental entities like the national ministry of health as well as academic institutions and individual citizens. To our knowledge, the overwhelming majority of these dashboards display raw data together with descriptive statistics of "hard" facts and numbers on COVID-19. To fill a gap, we build a dashboard with processed data from three different sources to track the sentiment in Austrian social media during COVID-19. It is easily accessible online and updated on a daily basis to give feedback to authorities and the interested general public.

\section{Method}

\begin{figure}[h!]
\begin{center}
\includegraphics[height=0.9\textheight]{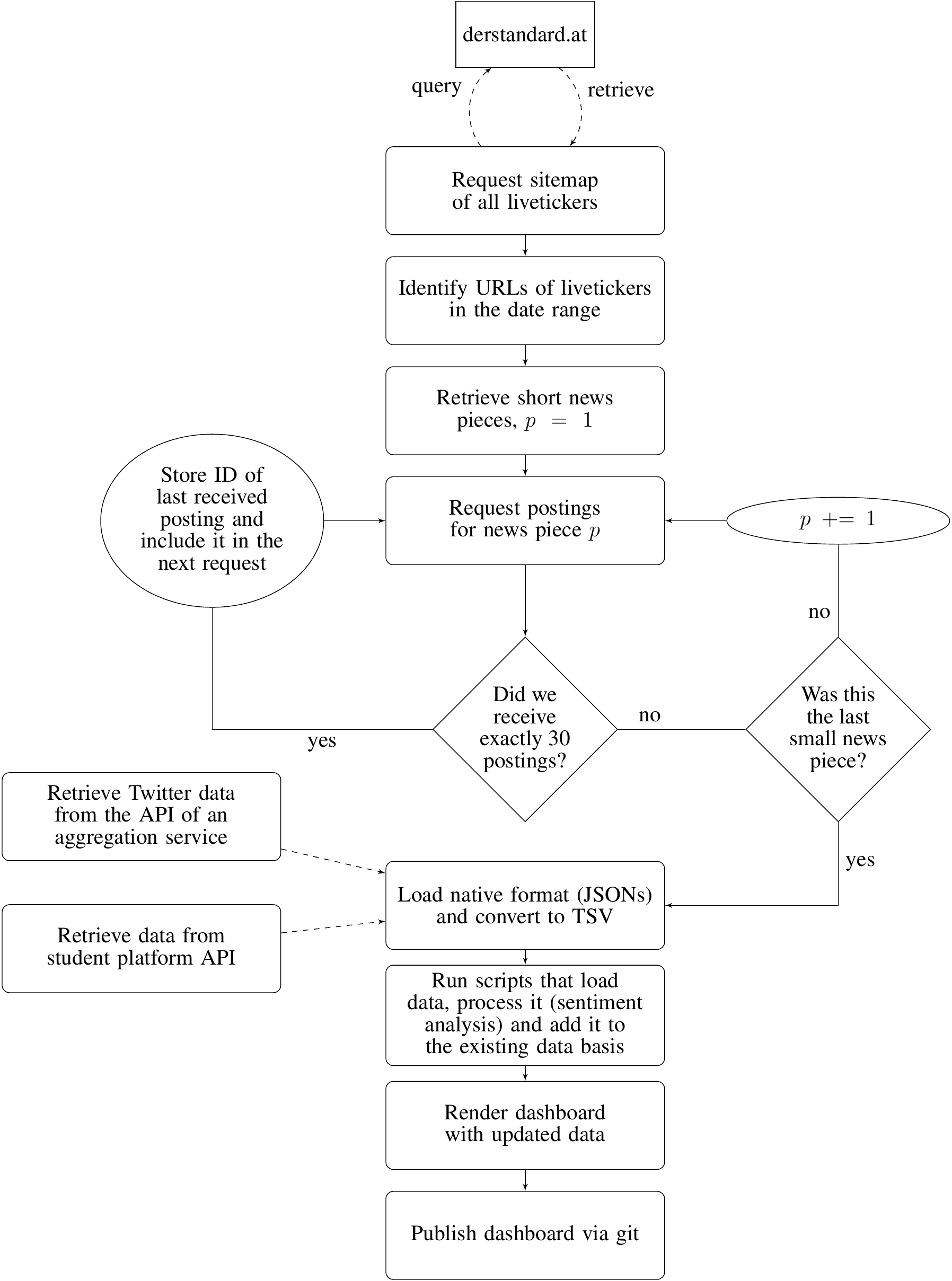}
\end{center}
\caption{\textbf{Flowchart of the daily routine of updating the dashboard.} We run this routine as a cronjob each day in the morning at 7am.}\label{fig:flowchart}
\end{figure}

We retrieve data from three different sites: a news platform, Twitter and a chat platform for students. All data for this article was gathered in compliance with the terms and conditions of the platforms involved. Twitter data was accessed through Crimson Hexagon (Brandwatch), an official Twitter partner. The platform for students and derstandard.at gave us their permission to retrieve the data automatically from their systems. A daily recurring task is set up on a server to retrieve and process the data, and to publish the updated page online (for a description of the workflow see Figure \ref{fig:flowchart}).

The news platform derstandard.at was an internet pioneer as it was the first German language newspaper to go online in 1995. From February 1999, it started entertaining an active community, first in a chatroom \cite{derstandard.atSTANDARDChatroomBar2018}. In 2008, the chatroom was converted to a forum that is still active today and allows for posting beneath articles. Users have to register to post and they can up- and down-vote single posts. In 2013 a platform change made voting more transparent by showing which user voted both positive or negative. According to a recent poll \cite{derstandard.atCoronaHoechstwerteFuer2020}, derstandard.at is considered both the most trustful and most useful source of information on COVID-19 in Austria. Visitors come from Austria, but also from other parts of the German-speaking area. In 2020, derstandard.at was visited by $2,546,000$ unique users per month that stay on average $06:42$ minutes on the site and request a total of $215,974,000$ subpages \cite{derstandard.atsalesteamDerstandardMediaData2020}. To cover the developments around COVID-19, daily livetickers (except Sundays) were set up on derstandard.at. Figure S1 in the Supplementary Information shows an example of the web interface of such a liveticker.

As no dedicated API exists for data retrieval from derstandard.at, we use web-scraping to retrieve the data (under permission from the site). First, we request a sitemap and identify the relevant URLs of livetickers. Second, we query each small news item of each of the livetickers. We receive data in JSON format and flatten and transform the JSON object to extract the ID of each small news piece. Third, we query the posts attached to that ID in batches. This is necessary because derstandard.at does not display all the posts at once beneath a small news item. Instead, the page loads a new batch of posts as soon as the user reaches the bottom of the screen. This strategy is chosen to not overcrowd the interface, as the maximum number of posts beneath one small news item can be very high (up to $2293$ posts in our data set). By following our iterative workflow to request posts, we are able to circumvent issues of pagination. Finally, after we have received all posts, we transform the JSON objects to tabulator-separated value files for further analysis. This approach is summarised in the upper part of Figure \ref{fig:flowchart}.

To retrieve daily values for our indicators from Twitter, we rely on the Forsight platform by \href{https://forsight.crimsonhexagon.com/}{Crimson Hexagon}, an aggregation service of data from various platforms, including Twitter. Twitter has an idiosyncratic user base in Austria, mainly composed of opinion makers, like journalists and politicians. In the case of studying responses to a pandemic, studying these populations gives us an insight into public sentiment due to their influence in public opinion. Yet, one should keep in mind that Twitter users are younger, more liberal, and have higher formal education than the general population \cite{pewresearchcenterHowTwitterUsers2019}.

As a third and last source, we include a discussion platform for young adults in Austria \footnote{Since the platform owners agreed to share data within the scope of the ongoing COVID-19 crisis but do not want to be named in public, we must refrain to give details that can identify the platform. However, it must be noted that all data analyzed here was publicly available and we did not analyze any private digital traces.}. The discussions on the platform are organized in channels based on locality, with an average of $580\pm390$ (mean $\pm$ standard deviation) posts per day from 2020-01-01 to 2020-05-27. The typical number of posts per day on the platform dropped from $830\pm260$ (January to April) to $160\pm80$ (April to May). This drop occurred due to the removal of the possibility to post anonymously on April 4$^\text{th}$ 2020 in order to prevent hate speech. Based on data from this platform, we study the reaction of the special community of young adults in different Austrian locations, with the majority of posts originating in Vienna (9\%), Graz (8\%) and other locations (83\%). 

To assess expressions of emotions and social processes, we match text in posts on all three platforms to word classes in the German version of the Linguistic Inquiry and Word Count (LIWC) Dictionary \cite{wolf_computergestutzte_2008}, including  anxiety, anger, sadness, positive emotions and social terms. LIWC is a standard methodology in psychology for text analysis that includes validated lexica in German.  It has been shown that LIWC, despite its simplicity, has an accuracy to classify emotions in tweets that is comparable to other state of the art tools in sentiment analysis benchmarks \cite{ribeiro2016sentibench}. Previous research has shown that LIWC, when applied to large-scale aggregates of tweets, has similar correlations with well-being measures as other, more advanced text analysis techniques \cite{quercia2012tracking,jaidka2020estimating}. Since within the scope of this study only text aggregates will be analysed, LIWC is an appropriate method and can be applied to all sorts of text data that is collected for the monitor. For the prosocial lexicon, we translated a list of prosocial terms used in previous research \cite{frimer_moral_2014}, including for example words related to helping, empathy, cooperating, sharing, volunteering, and donating. 

We adapt the dictionaries to the task at hand by excluding most obvious terms that can bias the analysis, as done in recent research validating Twitter word frequency data \cite{jaidka2020estimating}. Specifically, we cleaned the lists for (1) words which are likely more frequently used during the COVID-19 pandemic e.g. by news media and do not necessarily express an emotion (sadness: tot*; anger: toete*, töt*, töte*; positive: heilte, geheilt, heilt, heilte*, heilung; prosocial: Heilverfahren, Behandlung, Behandlungen, Dienstpflicht, Öffentlicher Dienst, and Digitale Dienste all matching Dienst*), (2) potential mismatches unrelated to the respective emotion (sadness: Harmonie/harmlos matching Harm*; positive: äußerst; prosocial: Dienstag matching Dienst*) (3) specific Austria-related terms like city names (sadness: Klagenfurt matching klagen*) or events (sadness: Misstrauensantrag matching miss*), and (4)  Twitter-related terms for the analysis of Tweets only (prosocial: teilen, teilt mit). 

For text from derstandard.at, we average the frequency of terms per post to take into account the varying lengths of posts. As Twitter has a strict character limit of $280$ characters per post, Crimson Hexagon provides the number of tweets containing at least one of the terms, based on which we calculate the proportion of such posts. Posts have a median length of $61$ characters in derstandard.at, $101$ characters in Twitter, and $51$ characters in the chat platform for young adults. To exclude periodic weekday effects, we correct for the daily baseline of our indicators by computing relative differences to mean daily baseline values. For derstandard.at data, the baseline is computed from all posts to derstandard.at articles in the year 2019. We use the main website articles for this instead of livetickers because during 2019, livetickers were mainly used to cover sport events (for an example see \href{https://www.derstandard.at/jetzt/livebericht/2000088339740/bundesliga-live-lask-sturm}{https://www.derstandard.at/jetzt/livebericht/2000088339740/bundesliga-live-lask-sturm}) or high-profile court cases \linebreak
(\href{https://www.derstandard.at/jetzt/livebericht/2000088169126/buwog-prozess-vermoegensverwalter-stinksauer-auf-meischberger}{https://www.derstandard.at/jetzt/livebericht/2000088169126/buwog-prozess-vermoegensverwalter-stinksauer-auf-meischberger}). Thereby, we choose a slightly different medium for our baselines to avoid having a topic bias in the baselines. Nonetheless, it comes from the same platform with the same layout and functionalities and an overlapping user base: $14422$ users ($75\%$ of total unique users in the livetickers) in our data set that are active at livetickers also post at normal articles. The speed of posting may differ slightly, because the article is typically posted in a final format, whereas small news pieces are added constantly in livetickers. For the other data sources, we correct by computing the baseline for the indicators from the start of period available to us (Twitter back to 2013-01-01, chat platform for young adults back to 2019-01-01) to January 2020.

Finally, we combine the processed data and render an interactive website. For this, we use "plotly" \cite{plotly}, "rflexdashboard" \cite{flexdashboard} and "wordcloud2" \cite{wordcloud2} in R \cite{R}, and the "git" protocol to upload the resulting HTML page to GitHub Pages. Using versioning control allows us to easily revert the page to a previous state in case of an error.

\section{Results}

\begin{figure}[h!]
\begin{center}
\includegraphics[width=1.2\textwidth]{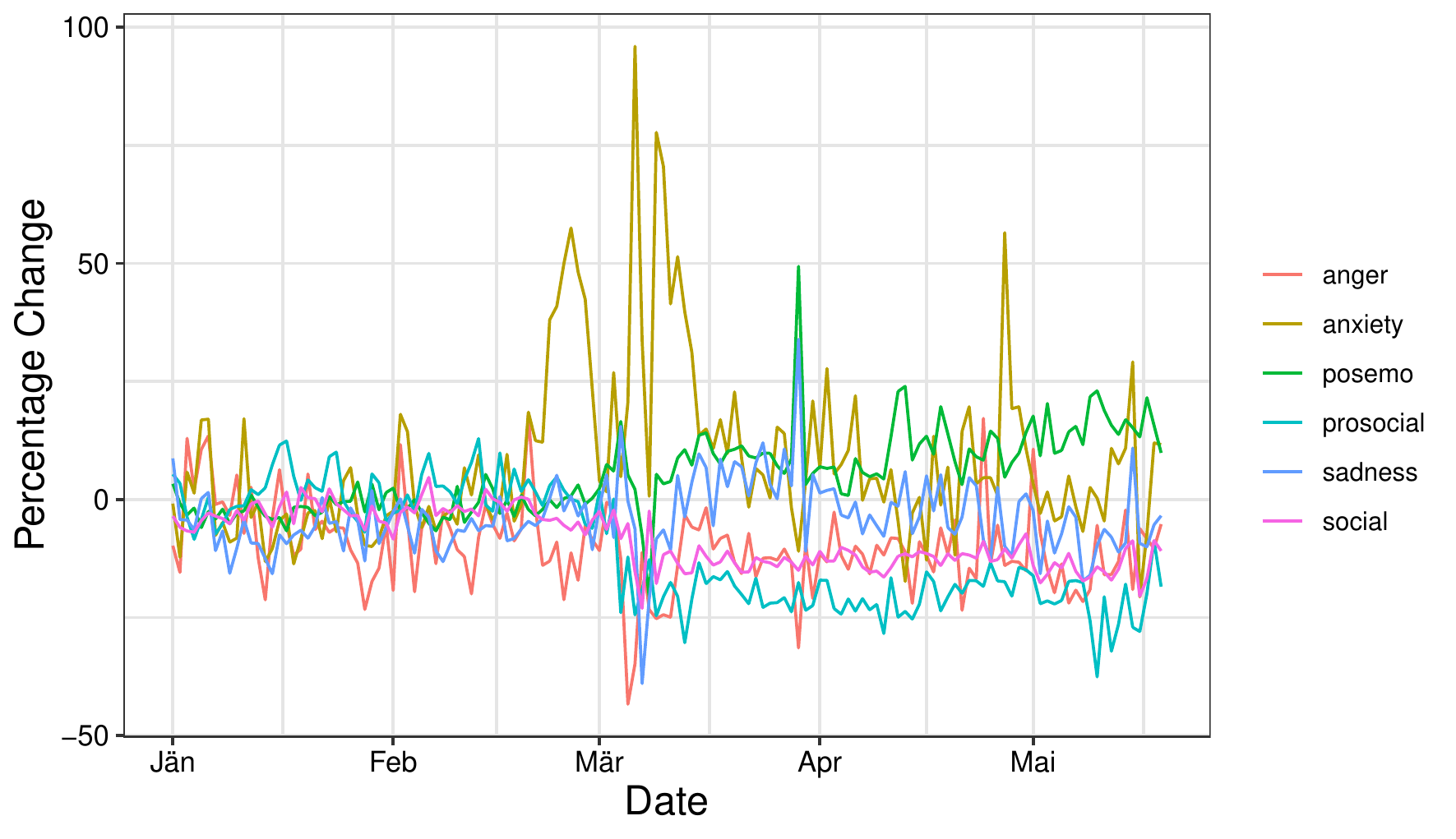}
\end{center}
\caption{\textbf{Timeline of the indicators for derstandard.at during the COVID-19 period.} Values correspond to the percentage change against the baseline of the full year 2019. To be included, dictionary terms have to appear at least 10 times in both corpora.}\label{fig:liveticker}
\end{figure}

We track the sentiment of the Austrian population during  COVID-19 and make our findings available as an interactive online dashboard that is updated daily. We display the time series almost in real-time with a small delay to catch all available data (see Figure \ref{fig:liveticker} using derstandard.at as a data source). It has features such as the option to display the number of observations by hovering over the data point or to isolate lines and to compare only a subset of indicators. The dashboard can be accessed online at \linebreak \href{http://www.mpellert.at/covid19_monitor_austria/}{http://www.mpellert.at/covid19\_monitor\_austria/}.

Table \ref{tab:nobs} shows several descriptive statistics of the data sets used. For derstandard.at, we retrieved $111$ livetickers with $10013$ small news items. On average, users publish $183\pm156$ posts under each of those items in the time period of interest (2020-03-09 to 2020-06-03). Posts have a median length of $61$ characters (see Figure S2 for a histogram of the length of posts). Posts provide immediate reactions by the users of derstandard.at: The median is at $24.7$ seconds for the first post to appear below a small news item.

\begin{figure}[h!]
\begin{center}
\includegraphics[width=0.82\textwidth]{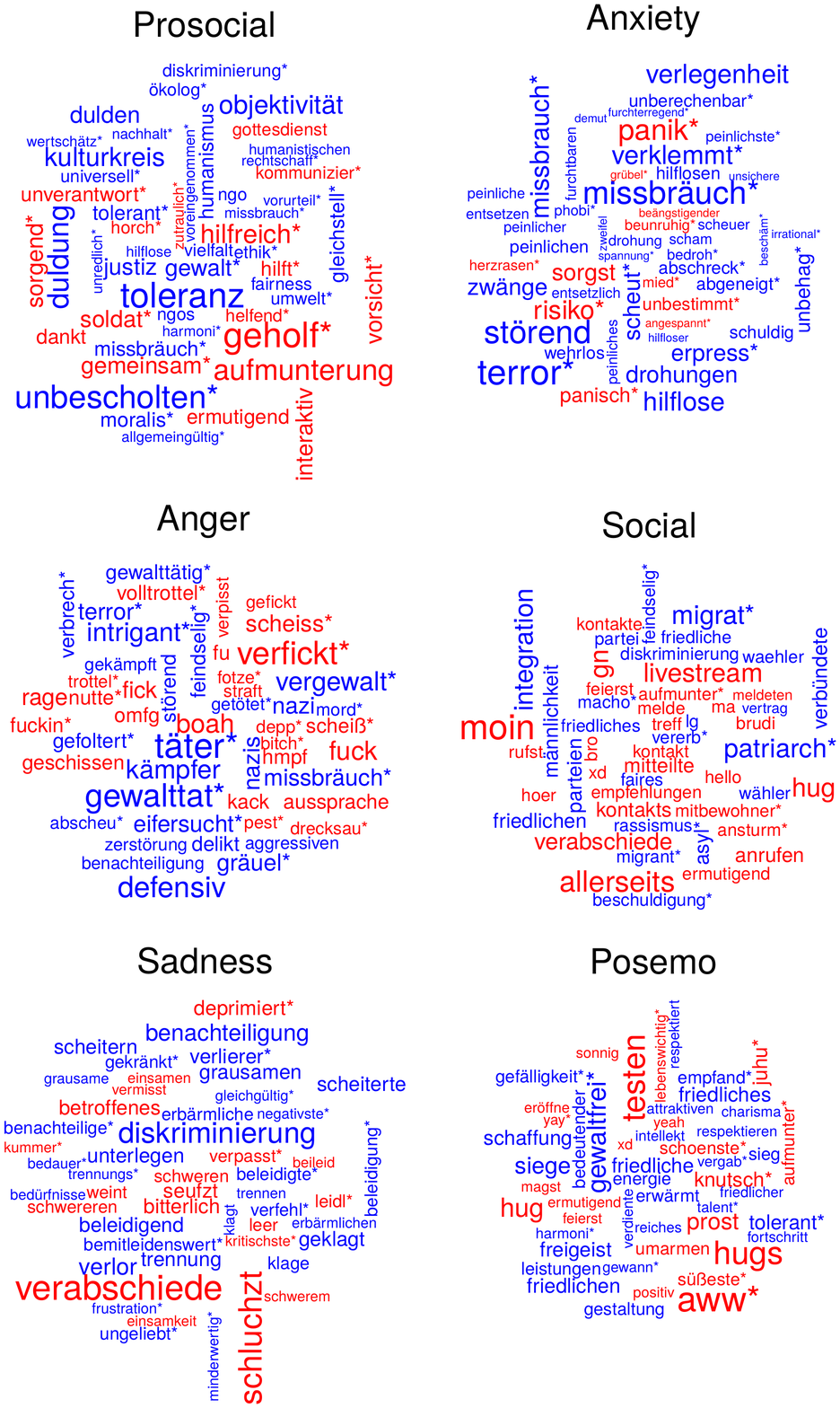}
\end{center}
\caption{\textbf{Wordclouds for posts on derstandard.at showing the matched words in each category.} Size corresponds to the magnitude and color to the direction of change: blue and red mean less and more prevalent in the COVID-19 livetickers than in the normal articles of 2019, respectively.  To be included, dictionary terms have to appear at least 10 times in both corpora.}\label{fig:wcs}
\end{figure}

We use word clouds (Figure \ref{fig:wcs}) to visualize the emotional content of posts. While livetickers on COVID-19 cover the time period from 2020-03-09 until 2020-06-03, the baseline includes normal articles on derstandard.at from 2019.  To highlight changes in language use during COVID-19, our word clouds compare word frequency in the livetickers with the baseline: The size of words in the clouds is proportional to $|\log(\frac{prob_{livetickers}}{prob_{baseline}})|$, where $prob_{baseline}$ and $prob_{livetickers}$ refer to the frequency of the dictionary term compared to the frequency of all matches of terms in that category, in the baseline and the livetickers, respectively. Color of words corresponds to the sign of this quantity: Red means positive, i.e. the frequency of the word increased in the livetickers, whereas blue signifies that the usage of the word decreased. By combining these information, our word clouds give an impression on how the composition of terms in the dictionary categories changed during COVID-19.

Our dashboard analyses a part of public discourse. We assume that the lockdown of public life increased tendencies of the population to move debates online. Users that take part in these discussions often form very active communities that sometimes structure their whole day around their posting activities. This is reflected in our data in the word clouds of Figure \ref{fig:wcs} from the increased usage of greetings (category "social"), marking the start or the end of a day such as "moin"/"good morning" or "gn"/"good night".

We identified the following events in Austria corresponding to anxiety spikes in expressed emotions in social media. Unrelated to COVID-19, there was reporting on a terrorist attack in Hanau, Germany on 2020-02-25. The first reported COVID-19 case in Austria was on 2020-02-25 and the first death on 2020-03-12. The first press conference, announcing bans of large public events and university closures as first measures, happened on 2020-03-10. It was followed by strict social distancing measures announced on 2020-03-15, starting on the day after. The overall patterns in the monitor of sentiment in Figure \ref{fig:liveticker} show that Austrian user's expressions of anxiety increased, whereas anger decreased in our observation period. We go into detail on this in Section \ref{sec:discussion}.

The sentiment dynamics on social media platforms can be influenced by content that spreads fear and other negative emotions. Timely online emotion monitoring could help to quickly identify such campaigns by malicious actors. Even legitimately elected governments can follow the controversial strategy of steering emotions to alert the population to the danger of a threat. For example, democratically elected actors can deliberately elicit emotions such as fear or anxiety to increase compliance from the top down. Such a strategy has been followed in Austria \cite{orf.atRegierungsprotokollAngstVor2020} and other countries like Germany \cite{AngstVorInfektionErwuenschtDE}. Reports about the deliberate stirring of fear by the Austrian government are reflected in a spike of anxiety on 2020-04-27 in Figure \ref{fig:liveticker}. The spikes of anxiety at the beginning of March in the early stages of the COVID-19 outbreak may have been reinforced by anxiety eliciting strategies.

In an effort to provide an archive of Austrian web resources for future reference, the Austrian National Library (ÖNB) monitors the dashboard and stores changes. There are a number of such initiatives also in other nations \cite{gomesSurveyWebArchiving2011} with the earliest and most famous example being \href{archive.org}{archive.org}. Through selective harvesting of resources connected to COVID-19, the dashboard is part of the ÖNB collection "Coronavirus 2020" (\href{https://webarchiv.onb.ac.at/}{https://webarchiv.onb.ac.at/}).

\section{Discussion} \label{sec:discussion}

Our results show patterns in the change of language use during COVID-19. In the anger category, words related to violence and crime are less frequent in livetickers since COVID-19 compared to 2019, indicating that reports and discussions about violent events, or possibly even these events themselves, become less frequent as the public discourse focuses on events related to the pandemic. For anxiety, the most remarkable change is a reduction in words related to terror and abuse, accompanied by a smaller increase of terms linked to panic, risk and uncertainty. In the sadness category, the verb "verabschiede"/"saying goodbye" appears almost 9 times more often in the livetickers. For prosocial words, terms referring to helping, community and encouragement increased. From the social terms, the word "empfehlungen"/"recommendations" occurs slightly more frequently, while topics of migration, integration and patriarchy are less often discussed. Finally, positive terms that increase the most are the expression of admiration "aww*" and "hugs", indicating that people send each other virtual hugs instead of physical ones.   

Dynamics of collective emotions may be different in crisis times. While they typically vary fast \cite{pellertIndividualDynamicsAffective2020} and return to the baseline within a matter of days even after catastrophic events like natural disasters or terrorist attacks \cite{gruebner_novel_2017, garcia_collective_2019}, changes during the COVID-19 pandemic in Austria have lasted several weeks for most analysed categories (up to 12 weeks in some cases). In contrast to one-off events, threat from a disease like COVID-19 is more diffuse, and the emotion-eliciting events are distributed in time. In addition, measures that strongly affect people's daily lives over a long period of time, as well as high level of uncertainty, likely contribute to the unprecedented changes of collective emotional expression in online social media.

The dashboard illustrates early and strong increases in anxiety across all three analysed platforms starting at the time of the first confirmed cases in Austria (end of February 2020). A first initial spike of anxiety-terms occurs on all three platforms around the time the first positive cases were confirmed and news about the serious situation in Italy were broadcast in Austria. About two weeks later, levels rise again together with the number of confirmed cases, reaching particularly high levels in the week before the lock-down on 16 March. Afterwards, the gradually drop again. In total, levels of anxiety-expression did not return to the baseline for more than six weeks from 2020-02-22 until 2020-04-07 on Twitter. On derstandard.at, levels also remained above the baseline for more than four weeks in a row. Timelines for Twitter and derstandard.at also show a clear and enduring decrease of anger-related words starting in the week before the lock-down, as discussions of potentially controversial topics other than COVID-19 become scarcer. This decrease lasts for four weeks on derstandard.at (21 Februar - 23 April), but is particularly stable on Twitter, where anger-terms remain less frequent than in 2019 for 2.5 months in a row (2020-03-09 to 2020-05-29). In contrast, prosocial and social terms show opposing trends on these two platforms: They increase slightly but do so for more than 2 months on Twitter, where people share not only news, but also talk about their personal lives. In contrast, they decrease  for more than 3 months in a row on derstandard.at, where people mostly discuss specific political events or topics.The increase of sadness-related expressions is smaller than changes in anxiety and anger, but also lasted for about a month on Twitter, and two weeks on derstandard.at. Interestingly, positive expressions were used slightly more frequently on all three platforms for long periods since the outbreak. This trend is visible from the beginning of March on the student platform and derstandard.at, and further increases since restrictions on people's lives have reduced. In total, positive expressions are more frequent than baseline during the last 2.5 months (as of 13$\text{th}$ of June) on derstandard.at. An analysis of collective emotions in Reddit comments from users in eight US cities found results similar to ours, including spikes in anxiety and the decrease in anger \cite{ashokkumarUnfoldingCOVIDOutbreak2020}, which suggests that some of our findings might generalize to other platforms and countries.

The dashboard gives opinion makers and the interested public a way to observe collective sentiment vis-a-vis the crisis response in the context of a pandemic. It has gained attention from Austrian media \cite{apaCoronavirusOnlineEmotionenWeniger2020}, and from the COVID19 Future Operations Clearing Board \cite{federal_chancellery_republic_of_austria_covid-19_2020}, an interdisciplinary platform for exchange and collaboration between researchers put in place by the Federal Chancellery of the Republic of Austria. Especially during the first weeks of the crisis, multiple newspapers reported on the changes of emotional expressions in online platforms \cite{apa_online-emotionen_2020, apa_coronavirus_2020, wiener_zeitung_online_gefuhle_2020, ennemoser_online-emotionen_2020, scienceorfat_online-emotionen_2020}. Timely knowledge about the collective emotional state and expressed social attitudes of the population is valuable for adapting emergency and risk-communication as well as for improving the preparedness of (mental) health services.  

\section*{Conflict of Interest Statement}
The authors declare that the research was conducted in the absence of any commercial or financial relationships that could be construed as a potential conflict of interest.

\section*{Author Contributions}
MP and DG designed research. MP retrieved derstandard.at data, processed and analyzed all data, and implemented the dashboard. JL retrieved data for the platform for young adults. HM retrieved data for Twitter, and wrote methods and result reports for the dashboard. MP, JL and HM wrote the draft of the manuscript. All authors provided input for writing and approved the final manuscript.

\section*{Funding}
This work was funded by the Vienna Science and Technology Fund through the project “Emotional Well-Being in the Digital Society” (Grant No. VRG16-005).

\section*{Acknowledgments}
We thank Christian Burger from derstandard.at for providing data access, and Julia Litofcenko and Lena Müller-Naendrup for their support in translating the prosocial dictionary to German. Access to Crimson Hexagon was provided via the project V!brant Emotional Health grant “Suicide Prevention media campaign Oregon” to Thomas Niederkrotenthaler. 

\section*{Supplemental Data}
Supplementary Material is included.

\section*{Data Availability Statement}
The dashboard can be accessed at \href{http://www.mpellert.at/covid19_monitor_austria/}{http://www.mpellert.at/covid19\_monitor\_austria/}. The source code is available at \href{https://github.com/maxpel/covid19_monitor_austria}{https://github.com/maxpel/covid19\_monitor\_austria}. The data sets accumulated daily by updating the dashboard will be released in the future.

\bibliographystyle{unsrt}
\bibliography{dashboard}

\clearpage{}

\section*{Figures and Tables}

\begin{table}[h!]
\centering
\begin{tabular}{llll}
\toprule
& derstandard.at & Twitter & student platform\\
\midrule
\# Posts & 1,827,576 & 2,001,420 & 158,022 \\
Mean \# Posts per day & 21,007 & 23,005 & 1,816\\
\# Unique Users & 19,263 & 594,500 & NA \\
Fraction posemo & 0.333 & 0.435 & 0.383\\
Fraction anxiety & 0.036 & 0.031 & 0.032\\
Fraction anger & 0.062 & 0.057 & 0.062\\
Fraction sad & 0.083 & 0.064 & 0.09\\
Fraction social & 0.561 & 0.559 & 0.592\\
Fraction prosocial & 0.105 & 0.191  & 0.099\\
\bottomrule
\end{tabular}
\caption{\textbf{Descriptive statistics showing relevant aspects of the data sources.} Numbers refer to the time period from 9 March to 3 June 2020 (87 days). The total number of Twitter users in Austria in January 2019 is taken from the report of DataReportal \cite{datareportal_digital_2019}. Fractions refer to the number of posts containing at least one term from the relevant dictionary category in LIWC divided by the total number of posts.}\label{tab:nobs}
\end{table}

\end{document}


\author[$\dagger$,$\star$]{Max Pellert}
\author[$\dagger$,$\star$]{Jana Lasser}
\author[$\dagger$,$\star$,$\ddag$]{Hannah Metzler}
\author[$\dagger$,$\star$]{David Garcia}

\affil[$\dagger$]{Complexity Science Hub Vienna, Vienna, Austria}
\affil[$\star$]{Section for Science of Complex Systems, Center for Medical Statistics, Informatics and Intelligent Systems, Medical University of Vienna, Vienna, Austria}
\affil[$\ddag$]{Institute for Globally Distributed Open Research and Education}

\title{Supplementary Material of Dashboard of sentiment in Austrian social media during COVID-19}

\maketitle

\clearpage
\section{derstandard.at Interface}

\begin{figure}[h!]
\begin{center}
\includegraphics[width=0.95\textwidth]{liveticker_whole_cropped_rectangles.pdf}
\end{center}
\caption{\textbf{Example of derstandard.at liveticker} The rectangle in navy blue shows one small news item. New postings can be added using the form to the right of it (fuchsia rectangle) and the current total number of postings in the liveticker is shown. Below, the postings connected to the news item are displayed (maroon rectangle). \label{fig:liveticker}}
\end{figure}

\clearpage
\section{Length of postings at derstandard.at}

\begin{figure}[h!]
\begin{center}
\includegraphics[width=0.95\textwidth]{no_character.pdf}
\end{center}
\caption{\textbf{Histogram of the number of characters in derstandard.at postings with a log-transformed x-axis. }\label{fig:nochar}}
\end{figure}